\documentclass[prb,preprint,showpacs]{revtex4}
\usepackage{amsmath}
\usepackage{amsfonts}
\usepackage{amssymb}
\usepackage{array}
\usepackage{floatflt}
\usepackage{graphicx}
\usepackage{longtable}

\begin{document}

\title{Local Coarse-grained Approximation to Path Integral Monte Carlo Integration for Fermion Systems}
\author{D. Y. Sun}
\address{State Key Laboratory of Precision Spectroscopy and Department of Physics, East China Normal
University, Shanghai 200062, China}

\begin{abstract}
An approximate treatment of exchange in finite-temperature path
integral Monte Carlo simulations for fermions has been proposed. In
this method, some of the fine details of density matrix due to
permutations have been smoothed over or averaged out by using the
coarse-grained approximation. The practical usefulness of the method
is tested for interacting fermions in a three dimensional harmonic
well. The results show that, the present method not only reduces the
sign fluctuation of the density matrix, but also avoid the fermion
system collapsing into boson system at low temperatures. The method
is substantiated to be exact when applied to free particles.

\end{abstract}

\maketitle

\section{Introduction}

The path integral Monte Carlo method (PIMC) provides a
nonperturbative, basis-set-independent, and fully correlated
calculation for quantum many-body systems at both zero and finite
temperature.\cite{Ceperley95,Foulkes01,sign} However for
many-fermion systems, PIMC suffers from uncontrollable errors
arising from the notorious \emph{sign problem},\cite{sign,Loh90}
which limits the accuracy or stability of the method. The origin of
the \emph{sign problem} comes from the fact that the density matrix
can be positive or negative by even or odd permutations. At low
temperatures, contributions from positive and negative parts of the
density matrix almost perfectly cancel each other so that there is
no hope of extracting any useful information.

A few methods have been proposed to deal with the \emph{sign
problem}.  For problems in the continuous space, there are
fixed-node approximation,\cite{Anderson76,Ceperley92} released node
methods,\cite{Ceperley80} exact cancelation methods with Green's
function sampling,\cite{Chen95} multilevel blocking
algorithm,\cite{Mak98,Egger00} a hybrid path integral and basis set
method,\cite{Chiles84} various pseudopotential
approximations,\cite{Schnitker87,Landman87,Hall88,Kuki87,Coker87,Bartholomew85,Sprik85,Oh98,Miura01}
the general method for replacing integration over pure states by
integration over idempotent density matrices,\cite{Newman91} and
method by introducing several images of the system,\cite{Lyubartsev}
global stationary phase approach.\cite{Moreira} There are also a
number of methods for lattice
models.\cite{Zhang97,Zhang99,Helenius00} So far many efforts have
been devoted to tackle this problem, however, it remains being the
key bottleneck in using PIMC for many-fermion systems.

 In this paper, we use a
methodology to reduce the rapid oscillation of integrand in the
evaluation of high dimensional integrals. The idea is that, for the
region in which the function is rapidly oscillating, the
coarse-grained approximations are used to kill fluctuations.
Applying this technique to PIMC, we found that, the sign
fluctuations of density matrix can be reduced, thus the Metropolis
MC integration algorithm converges efficiently. More importantly,
after carrying out this approximation, the exchange determinant
becomes a nonlocal form in imaginary time, thus the collapse
behavior can be avoided (see below). The basic strategy can also be
used in the evaluation of other integrals, where the integrands
exhibit the rapidly oscillating characters.

This paper is organized as follows. The methodology is described in
Sec. II. The numerical tests are presented in Sec. III. Discussion
and conclusion are given in Sec. IV.

\section{Methodology}

To illustrate the coarse-grained approximation used in present
paper, we first consider the following integral,
\begin{equation}
I=\int_{a}^{c}\int_{A}^{C}f(x,y)g(x,y)h(x,y)dxdy.
\end{equation}
We assume that, except for \emph{f(x,y)} is a rapidly oscillating
function for variable \emph{y}, the rest parts of integrands are
well behavior (or slow varied) function of \emph{x} and \emph{y}.
Due to the rapid oscillation of  \emph{f(x,y)}, it could cause the
difficulty on evaluating the integral (Eq. 1) in MC simulation. To
overcome the difficulty, our strategy is to make coarse-grained
approximations to \emph{f(x,y)}. To do this, we rewrite the above
integral as,
\begin{equation}
I=\int_{a}^{c}\int_{A}^{C}F(x)g(x,y)h(x,y)dxdy,
\end{equation}
with
\begin{equation}
F(x)=\frac{\int_{A}^{C}f(x,y')g(x,y')h(x,y')dy'}{\int_{A}^{C}g(x,y')h(x,y')dy'}.
\label{eq:partation1}
\end{equation}
 One can see that, \emph{F(x)} is a kind of
coarse-grained functions, and the rapid oscillation of \emph{f(x,y)}
is smoothed over. \emph{F(x)} also can be viewed as an average of
\emph{f(x,y)} weighted by \emph{g(x,y)}\emph{h(x,y)}. If \emph{F(x)}
can be evaluated either exactly or approximately, the rapid
fluctuation due to \emph{f(x,y)} could be reduced effectively. For
real problems, \emph{F(x)} usually is hard to be evaluated exactly.
However, it is possible to determine \emph{F(x)} under some
reasonable approximations, as we have done in present paper.

Considering a three-dimensional system consisting of
\emph{N} spinless, indistinguishable quantum fermions, the standard
PIMC is based on the following expansion of partition function:

\begin{equation}
Z=\frac{C}{N!}\lim_{M\rightarrow\infty}\int\prod_{i=1}^{N}\prod_{\nu=1}^{M}d\vec{r}_{i}^{(\nu)}detAexp(-\beta
H) \label{eq:partation1}
\end{equation}
with
\[H=\sum_{i=1}^{N}\frac{1}{2}m\omega_{M}^{2}L_{i}^{2}+\sum_{\nu=1}^{M}\frac{1}{M}V(\{\vec{r}_{i}^{(\nu)}\})\]
where $\omega_{M}=\frac{\sqrt{M}}{\beta\hbar}$,
$C=(\frac{mM}{2\pi\beta\hbar^{2}})^{\frac{3}{2}NM}$,
$\vec{r}_{i}^{(M+1)}=\vec{r}_{i}^{(1)}$, \emph{V} is the potential
energy, and the square \emph{length}($L_{i}^{2}$) is defined as,
$L_{i}^{2}=\sum_{\nu=1}^{M}(\vec{r}_{i}^{(\nu+1)}-\vec{r}_{i}^{(\nu)})^{2}$.
The subscript \emph{i} refers to the particle number while the
superscript $\nu$ refers to different slit of imaginary time.
$\beta$, $m$ and $M$ are reciprocal temperature($1/k_{B}T$), mass of
particles and total number of beads respectively.
$\{\vec{r}_{i}^{(\nu)}\}$ refers to
$(\vec{r}_{1}^{(\nu)},\vec{r}_{2}^{(\nu)},......,\vec{r}_{N-1}^{(\nu)},\vec{r}_{N}^{(\nu)})$.To
make the expression compact, we have introduced a $N\times N$ matrix
\emph{A} whose element reads
\begin{equation}
A_{ij}=exp(-\frac{1}{2}\beta
m\omega^{2}_{M}((\vec{r}_{i}^{(1)}-\vec{r}_{j}^{(M)})^{2}-(\vec{r}_{i}^{(1)}-\vec{r}_{i}^{(M)})^{2})).
\end{equation}
\emph{detA} is the determinant of the matrix \emph{A}, which
accounts the contribution of permutations to the partition function.
It is \emph{detA}, which can be positive or negative, that causes
the so-called \emph{sign problem}. Previous studies based on
pseudopotential methods have shown that, direct use of Eq.4-like
formula usually results in a fermion system \emph{collapsing} into a
bosonic state at low temperature.\cite{Hall88} The physical reason
comes from the fact that the matrix \emph{A} approaching to unit at
low temperature. To prevent this undesirable behavior, people
usually recast matrix \emph{A} in a nonlocal form as suggested by
Hall,\cite{Hall88} or directly use a nonlocal pseudopotential as
suggested by Miura and Okazaki.\cite{Miura01} Although these schemes
do give a good solution, the computational cost also increases.

From Eq. 5, one can see that, if $\vec{r}_{i}^{(M)}$ is close to
$\vec{r}_{j}^{(M)}$, $A_{ij}$ will be a rapidly oscillating function
of $\vec{r}_{i}^{(1)}$. Although, for N$>$2, it is difficult to
prove the direct relation between the rapid oscillation of $A_{ij}$
and the sign problem, it is quite clear that the rapid oscillation
of $A_{ij}$ directly results in the sign fluctuation for N=2.  To
smooth the rapid oscillation, the coarse-grained approximation is
made for $A_{ij}$ by integration over $\vec{r}_{i}^{(1)}$ for all
possible configurations with fixed $\vec{r}_{i}^{(M)}$,
$\vec{r}_{j}^{(M)}$ and $L_{i}$. Now Eq. 4 is replaced by

\begin{equation}
Z=\frac{C}{N!}\lim_{M\rightarrow\infty}\int\prod_{i=1}^{N}\prod_{\nu=1}^{M}d\vec{r}_{i}^{(\nu)}det\Xi
exp(-\beta H) \label{eq:partation1}
\end{equation}
 where the new $N\times N$ matrix $\Xi$ is the coarse-grained
approximation of matrix \emph{A}.  According to the idea presented
in Eq. 1, 2 and 3, the elements of $\Xi$ are
\begin{equation}
\Xi_{\alpha\gamma}=\frac{\int'\prod_{\nu=1}^{M-1}d\vec{r}_{\alpha}^{(\nu)}
A_{\alpha\gamma}exp(-\beta H)}
{\int'\prod_{\nu=1}^{M-1}d\vec{r}_{\alpha}^{(\nu)}exp(-\beta H)}
\end{equation}
Since $A_{\alpha\gamma}$ is only relevant to
$\vec{r}_{\alpha}^{(1)}$, $\vec{r}_{\alpha}^{(M)}$ and
$\vec{r}_{\gamma}^{(M)}$, Eq. 7 can be rewritten as,
\begin{equation}
\Xi_{\alpha\gamma}=\frac{\int'\prod_{\nu=1}^{M-1}d\vec{r}_{\alpha}^{(\nu)}
A_{\alpha\gamma}exp(-\frac{1}{2}\beta
m\omega_{M}^{2}L_{\alpha}^{2}-\beta U)}
{\int'\prod_{\nu=1}^{M-1}d\vec{r}_{\alpha}^{(\nu)}exp(-\frac{1}{2}\beta
m\omega_{M}^{2}L_{\alpha}^{2}-\beta U)}
\end{equation}
with
\[
U=\sum_{\nu=1}^{M}\frac{1}{M}V(\{\vec{r}_{i}^{(\nu)}\}))
\]
$\int'$ in Eq. 7 and 8 refers the integral under the constraint of
fixed $\vec{r}_{\alpha}^{(M)}$, $\vec{r}_{\gamma}^{(M)}$ and
$L_{\alpha}$.

Since the kinetic energy relevant
part($\frac{1}{2}\omega_{M}^{2}L_{\alpha}^{2}$) is a constant for
fixed $L_{\alpha}$. The Eq. 8 can be further simplified as,
\begin{equation}
\Xi_{\alpha\gamma}=\frac{\int'\prod_{\nu=1}^{M-1}d\vec{r}_{\alpha}^{(\nu)}
A_{\alpha\gamma}exp(-\beta U)}
{\int'\prod_{\nu=1}^{M-1}d\vec{r}_{\alpha}^{(\nu)}exp(-\beta U)}
\end{equation}
 To calculate the Eq. 9 under the constraint of
fixed $L_{\alpha}$, we can rewrite it as,

\begin{equation}
\Xi_{\alpha\gamma}=\frac{\int d\vec{r}_{\alpha}^{(1)}
A_{\alpha\gamma}\Delta(L_{\alpha},\vec{r}_{\alpha}^{(1)},
\vec{r}_{\alpha}^{(M)})\bar{U}} {\int
d\vec{r}_{i}^{(1)}\Delta(L_{\alpha},\vec{r}_{\alpha}^{(1)},
\vec{r}_{\alpha}^{(M)})\bar{U}}
\end{equation}
where $\Delta(L_{\alpha},\vec{r}_{\alpha}^{(1)},
\vec{r}_{\alpha}^{(M)})=\int'\prod_{\nu=2}^{M-1}d\vec{r}_{\alpha}^{(\nu)}$,
which is the number of configurations for fixed $L_{\alpha}$,
$\vec{r}_{\alpha}^{(1)}$ and $\vec{r}_{\alpha}^{(M)}$. $\bar{U}$
reads,
\begin{equation}
\bar{U}=\int\prod_{\nu=2}^{M-1}d\vec{r}_{\alpha}^{(\nu)}exp(-\beta
U)/\Delta(L_{\alpha},\vec{r}_{\alpha}^{(1)},\vec{r}_{\alpha}^{(M)})
\end{equation}
 The above integral is also under the constraint of fixed $L_{\alpha}$. There is almost no hope to evaluate Eq.
 10
exactly. In this paper, we calculate it with approximations,

\begin{equation}
\Xi_{\alpha\gamma}\approx\frac{1}{\Upsilon(L_{\alpha},\vec{r}_{\alpha}^{(M)})}\int
d\vec{r}_{\alpha}^{(1)}A_{\alpha\gamma}\Delta(L_{\alpha},\vec{r}_{\alpha}^{(1)},
\vec{r}_{\alpha}^{(M)}), \label{eq:partation3}
\end{equation}
 $\Upsilon(L_{\alpha},\vec{r}_{\alpha}^{(M)})$ is the
total number of configurations for fixed $L_{\alpha}$ and
$\vec{r}_{\alpha}^{(M)}$, \emph{i.e},
\[\Upsilon(L_{\alpha},\vec{r}_{\alpha}^{(M)})=\int d\vec{r}_{\alpha}^{(1)}\Delta(L_{\alpha},\vec{r}_{\alpha}^{(1)},\vec{r}_{\alpha}^{(M)}).\]

By replacing Eq. 10 with Eq. 12, we have assumed that $\bar{U}$ is
weakly dependent of $\vec{r}_{\alpha}^{(1)}$ for fixed
$\vec{r}_{\alpha}^{(M)}$, $\vec{r}_{\gamma}^{(M)}$ and $L_{\alpha}$.
This approximation works well if \emph{M} is not too small. The
reason lies on the fact that, only the configurations, in which
$\vec{r}_{\alpha}^{(1)}$ is close to $\vec{r}_{\alpha}^{(M)}$, make
the  $\Delta(L_{i},\vec{r}_{i}^{(1)},\vec{r}_{i}^{(M)})$ be
significant(see below Eq. 13 and 14). Our numerical test (see below)
also demonstrates this point.

$\Delta(L_{\alpha},\vec{r}_{\alpha}^{(1)}, \vec{r}_{\alpha}^{(M)})$
can be written as an integral over three Cartesian directions,
\begin{equation}
\Delta(L_{\alpha},\vec{r}_{\alpha}^{(1)},
\vec{r}_{\alpha}^{(M)})=\int\bar{\Delta}(L_{\alpha
x},x_{\alpha}^{(1)}, x_{\alpha}^{(M)})\bar{\Delta}(L_{\alpha
y},y_{\alpha}^{(1)}, y_{\alpha}^{(M)})\bar{\Delta}(L_{\alpha
z},z_{\alpha}^{(1)}, z_{\alpha}^{(M)})dL_{\alpha x}dL_{\alpha
y}dL_{\alpha z},
\end{equation}
where the integral is evaluated under the constraint:
$L_{\alpha}^{2}=L_{\alpha x}^{2}+L_{\alpha y}^{2}+L_{\alpha z}^{2}$.
$\bar{\Delta}(L_{\alpha x},x_{\alpha}^{(1)},x_{\alpha}^{(M)})$ is
the number of configurations for fixed $L_{\alpha x}$ ($L_{\alpha
x}^{2}=\sum_{\nu=1}^{M}(x_{\alpha}^{(\nu+1)}-x_{\alpha}^{(\nu)})^{2}$)
,$x_{\alpha}^{(1)}$ and $x_{\alpha}^{(M)}$. $\bar{\Delta}(L_{\alpha
y},y_{\alpha}^{(1)},y_{\alpha}^{(M)})$ and $\bar{\Delta}(L_{\alpha
z},z_{\alpha}^{(1)},z_{\alpha}^{(M)})$ are the counterparts of
$\bar{\Delta}(L_{\alpha x},x_{\alpha}^{(1)},x_{\alpha}^{(M)})$ along
\emph{y} and \emph{z} direction respectively.

To calculate $\bar{\Delta}(L_{\alpha
x},x_{\alpha}^{(1)},x_{\alpha}^{(M)})$, we define a
(M-1)-dimensional vector $\vec{R}$, of which Cartesian components
are
$(x_{\alpha}^{(1)}-x_{\alpha}^{(2)},....,x_{\alpha}^{(M-1)}-x_{\alpha}^{(M)})$.
First, for a given $L_{\alpha x}$, it requires
$|\vec{R}|$=$\sqrt{L_{\alpha
x}^{2}-(x_{\alpha}^{(1)}-x_{\alpha}^{(M)})^{2}}$, all the
configurations satisfying this condition lie on a surface of
(M-1)-dimensional super-sphere with radius equal to $\sqrt{L_{\alpha
x}^{2}-(x_{\alpha}^{(1)}-x_{\alpha}^{(M)})^{2}}$; Second, since the
Cartesian components of $\vec{R}$ are not independent, \emph{i.e},
the project of $\vec{R}$ on the (M-1)-dimensional unit vector is
$\frac{(x_{\alpha}^{(1)}-x_{\alpha}^{(M)})}{\sqrt{M-1}}$, this
condition defines a (M-1)-dimensional super-plane; Thus, all the
(M-1)-dimensional points, which attribute to $\bar{\Delta}(L_{\alpha
x},x_{\alpha}^{(1)},x_{\alpha}^{(M)})$,
 lie on a (M-2)-dimensional
super-spherical surface intersected by the (M-1)-dimensional
super-sphere and the (M-1)-dimensional super-plane. According to
analytic geometry in high dimensional space, $\bar{\Delta}(L_{\alpha
x},x_{\alpha}^{(1)},x_{\alpha}^{(M)})$ is the proportional area of
the (M-2)-dimensional super-spherical surface with radius equal to
$(R^{2}-\frac{(x_{\alpha}^{(1)}-x_{\alpha}^{(M)})^{2}}{(M-1)})^\frac{1}{2}$.
We end up with:
\[\bar{\Delta}(L_{\alpha x},x_{\alpha}^{(1)},
x_{\alpha}^{(M)})dRdx_{i}^{(1)}\propto
C_{\bar{\Delta}}(R^{2}-\frac{(x_{\alpha}^{(1)}-x_{\alpha}^{(M)})^{2}}{(M-1)})^\frac{M-3}{2}dRdx_{\alpha}^{(1)}\]
with
$C_{\bar{\Delta}}=(M-2)\pi^{\frac{M-2}{2}}/((M-2)/2)!$.\cite{StaMec}
By changing integration variable from ($R$, $x_{\alpha}^{1}$) to
($L_{\alpha x}$, $x_{\alpha}^{1}$), we have,

\begin{equation}
\bar{\Delta}(L_{\alpha x},x_{\alpha}^{(1)},
x_{\alpha}^{(M)})dL_{\alpha x}dx_{\alpha}^{(1)}\propto
C_{\bar{\Delta}}L_{\alpha x}(L_{\alpha
x}^{2}-(x_{\alpha}^{(1)}-x_{\alpha}^{(M)})^{2}\frac{M}{M-1})^{\frac{M-4}{2}}dL_{\alpha
x}dx_{\alpha}^{(1)}, \label{eq:partation5}
\end{equation}

Similarly, we can obtain $\bar{\Delta}(L_{\alpha
y},y_{\alpha}^{(1)}, y_{\alpha}^{(M)})$ and $\bar{\Delta}(L_{\alpha
z},z_{\alpha}^{(1)}, z_{\alpha}^{(M)})$, which have the same formula
as $\bar{\Delta}(L_{\alpha x},x_{\alpha}^{(1)}, x_{\alpha}^{(M)})$.
Substituting $\bar{\Delta}(L_{\alpha x},x_{\alpha}^{(1)},
x_{\alpha}^{(M)})$ in Eq. 14 for $\bar{\Delta}(L_{\alpha
x},x_{\alpha}^{(1)}, x_{\alpha}^{(M)})$ in Eq. 13, as well as
replacing the counterparts of $\bar{\Delta}(L_{\alpha
y},y_{\alpha}^{(1)}, y_{\alpha}^{(M)})$ and $\bar{\Delta}(L_{\alpha
z},z_{\alpha}^{(1)}, z_{\alpha}^{(M)})$ in Eq. 13,
$\Delta(L_{\alpha},\vec{r}_{\alpha}^{(1)}, \vec{r}_{\alpha}^{(M)})$
can be obtained. As a result, $\Xi_{\alpha\gamma}$ ought to be
calculated numerically.

One can see that, $\bar{\Delta}(L_{\alpha x},x_{\alpha}^{(1)},
x_{\alpha}^{(M)})$ quickly decays as a function of
$|x_{\alpha}^{(1)}- x_{\alpha}^{(M)}|$. The behaviors of
$\bar{\Delta}(L_{\alpha y},y_{\alpha}^{(1)}, y_{\alpha}^{(M)})$ and
$\bar{\Delta}(L_{\alpha z},z_{\alpha}^{(1)}, z_{\alpha}^{(M)})$ are
the same as that of $\bar{\Delta}(L_{\alpha x},x_{\alpha}^{(1)},
x_{\alpha}^{(M)})$. Accordingly,
$\Delta(L_{\alpha},\vec{r}_{\alpha}^{(1)},\vec{r}_{\alpha}^{(M)})$
also quickly decays as a function of $|\vec{r}_{\alpha}^{(1)}-
\vec{r}_{\alpha}^{(M)}|$. By employing the change of variables as
$\vec{r}_{\alpha}^{(1)}=\vec{r}_{\alpha}^{(M)}+\vec{\delta}_{\alpha}$,
and making coordinate transformation in spherical coordinates, we
can see that $\Xi_{\alpha\gamma}$ is a function of $L_{\alpha}$ and
$r_{\alpha\gamma}^{(M)}$
($r_{\alpha\gamma}^{(M)}=|\vec{r}_{\alpha}^{(M)}-\vec{r}_{\alpha}^{(M)}|$).

After carrying out above coarse-grained approximation, we find that,
the off-diagonal element $\Xi_{\alpha\gamma}$ is a function of both
$r_{\alpha\gamma}^{(M)}$ and $L_{\alpha}$, which has an explicit
nonlocal form. In contrast, the off-diagonal element
$A_{\alpha\gamma}$ is a local function in imaginary time. More
importantly, the off-diagonal element $A_{\alpha\gamma}$ could be
much larger or much smaller than 1.0, while the off-diagonal element
$\Xi_{\alpha\gamma}$ is less than 1.0 when $L_{\alpha}$ is not too
long (see Fig. 1). Thus, after replacing $A_{\alpha\gamma}$ with
$\Xi_{\alpha\gamma}$, at least for two-particle system, a lot of
sign fluctuations are well canceled. However, since the length of
path could be much longer at low temperature, the off-diagonal
element $\Xi_{\alpha\gamma}$ can be larger than 1.0 (see Fig. 1),
which will result in the presence of negative $det\Xi$. To further
reduce the negative sign, we have made the second stage of
coarse-grained approximation for longer paths. Using the similar
idea as above, we can replace $\Xi$ by a $N\times N$ matrix
$\Lambda$,

 \begin{eqnarray}
  \Lambda_{\alpha\gamma}&=\left\{\begin{array}{ll}
\Xi_{\alpha\gamma}, for L_{\alpha}\leq L_{\alpha}^{\ast}\\
\eta(r_{\alpha\gamma}^{(M)}), for L_{\alpha}>L_{\alpha}^{\ast}\\
                \end{array} \right.
 \end{eqnarray}
$L_{\alpha}^{\ast}$ is a function of $r_{\alpha\gamma}^{(M)}$, which
is determined by,
\begin{equation}
\eta(r_{\alpha\gamma}^{(M)})=\frac{\int_{L_{\alpha}^{\ast}}^{\infty}\prod_{\nu=1}^{M-1}D\vec{r}_{\alpha}^{(\nu)}
\Xi_{\alpha\gamma}exp(-\beta H)}
{\int_{L_{\alpha}^{\ast}}^{\infty}\prod_{\nu=1}^{M-1}D\vec{r}_{\alpha}^{(\nu)}exp(-\beta
H)}
\end{equation}

The above integral is made over all the configurations with
$L_{\alpha}>L_{\alpha}^{\ast}$. For an arbitrary interact potential,
the above integral is almost no hope to be evaluated exactly.
However it can be calculated with approximations,

\begin{equation}
\eta(r_{\alpha\gamma}^{(M)})\approx\frac{\int_{L_{\alpha}^{\ast}}^{\infty}
dL_{\alpha} \Xi_{\alpha\gamma} exp(-\frac{1}{2}\beta
m\omega_{M}^{2}L_{\alpha}^{2})\Upsilon(L_{\alpha},\vec{r}_{\alpha}^{(M)})}{\int_{L_{\alpha}^{\ast}}^{\infty}
dL_{\alpha} exp(-\frac{1}{2}\beta
m\omega_{M}^{2}L_{\alpha}^{2})\Upsilon(L_{\alpha},\vec{r}_{\alpha}^{(M)})}.
\end{equation}
In the evaluation of the above coarse-grained approximation, we have
assumed that total potential energy is constant when $L_{\alpha}$ is
longer than certain value, \emph{i.e},  $L_{\alpha}^{\ast}$. In
current work, we have taken $\eta(r_{\alpha\gamma}^{(M)})=1$ through
the whole paper.

Within the current approximations, $\Lambda$ has a few advantages
over the original matrix \emph{A}. First, our calculations have
shown that the off-diagonal element of $\Lambda$ is not larger than
1.0 anywhere (see Fig. 1), in constrast the off-diagonal element of
\emph{A} could be much larger than 1.0. Thus at least for
two-particle system, $\Lambda$ is always non-negative, the
\emph{sign problem} completely vanishes. Second, different from
\emph{A}, $\Lambda$ is nonlocal, which depends on the whole path.
The nonlocal behavior of $\Lambda$ can effectively avoid the
\emph{collapse} of fermion system into boson system at low
temperature. At low temperature, the length of path becomes longer
and longer, so the off-diagonal element of $\Lambda$ has more chance
being 1.0. This situation makes $\Lambda$ have little chance being
unit matrix. Our calculation also demonstrates this point. It should
be pointed out that the current formula is exact for free particles
(see APPENDIX).

Now we end up with the final formula for real calculations,
\begin{equation}
Z\cong
C\int\prod_{i=1}^{N}\prod_{\nu=1}^{M}D\vec{r}_{i}^{(\nu)}det\Lambda
exp[-\beta
\sum_{i=1}^{N}\sum_{\nu=1}^{M}\frac{1}{2}m\omega_{M}^{2}(\vec{r}_{i}^{(\nu+1)}-\vec{r}_{i}^{(\nu)})^{2}+
\sum_{\nu=1}^{M}\frac{1}{M}V(\{\vec{r}_{i}^{(\nu)}\})]
\label{eq:partation2}
\end{equation}
In real calculation, the element of $\Lambda$ is first numerically
integrated. At the same time, the derivative of $\Lambda$ respective
to temperature is also numerically calculated to account the
contribution to thermal energy. Eq. 18 can not be directly used in
standard MC, since for the fermionic systems, $det\Lambda$ is not
always positive. However Eq. 18 can be integrated using modified MC
technique, which is widely used previously.\cite{DeR81,Takahashi84}
To achieve this result, we first defined pseudo-Hamiltonian,
$H_{p}$, which is,
\[H_{p}=\sum_{i=1}^{N}\sum_{\nu=1}^{M}\frac{1}{2}m\omega_{M}^{2}(\vec{r}_{i}^{(\nu+1)}-\vec{r}_{i}^{(\nu)})^{2}+
\sum_{\nu=1}^{M}\frac{1}{M}V(\{\vec{r}_{i}^{(\nu)}\})+ln|det\Lambda|.\]
The thermodynamic average of a physical quantity Q is
\begin{equation}
\langle Q\rangle
=\frac{\int\prod_{i=1}^{N}\prod_{\nu=1}^{M}D\vec{r}_{i}^{(\nu)}Q(\{r_{i}^{(\nu)}\})sgn(det\Lambda)exp(-\beta
H_{p}\})}
{\int\prod_{i=1}^{N}\prod_{\nu=1}^{M}D\vec{r}_{i}^{(\nu)}sgn(det\Lambda)exp(-\beta
H_{p}\})\})},
\end{equation}
 where $sgn(det\Lambda)$ stands for the sign of $det\Lambda$ at a configuration.

It needs to be pointed out that, we have used the similar technique
 as most pseudopotential methods,\cite{Schnitker87,Landman87,Hall88,Kuki87,Coker87,Bartholomew85,Sprik85,Oh98,Miura01}
  but we do not recast matrix $\Lambda$ or extend $\Lambda$ into each imaginary
 time.

\section{The Numerical Tests}

 To illustrate the usefulness
of the current method, we have considered \emph{N} interacting
spinless fermions confined in a three-dimensional harmonic well,
which Hamiltonian reads,
\begin{equation}
H=\sum_{j=1}^{N}(\frac{\vec{p}_{j}^{2}}{2m}+\frac{m\omega^{2}}{2}\vec{r}_{j}^{2})+\sum_{i<j}^{N}V(r_{ij}),
\label{eq:partation5}
\end{equation}
where \emph{m}, $\vec{r}_{j}$, $\vec{p}_{j}$ and $V(r_{ij})$ are
mass, positions, momenta of the particles, and the inter-particle
interaction potential, respectively. For computational simplicity,
the units by which $m=\hbar=k_{B}=1$ are used through the rest of
this paper. In current calculations, we consider three cases,
\emph{i.e}, Case 1: $V(r_{ij})=0$, N=6 and $\omega^{2}=1$, no
interaction between particles, reflecting a standard harmonic
system; Case 2: N=6, $V(r_{ij})=-\frac{m\Omega^{2}}{2}r_{ij}^{2}$,
where the interaction is also harmonic one with
$\Omega^{2}=\frac{1.0}{4.0}$ and $\omega^{2}=4$; Case 3:
$V(r_{ij})=\frac{q^{2}}{|\vec{r}_{i}-\vec{r}_{j}|}$,
$\omega^{2}=0.320224986$, $q=1$ and N=2, interaction between
particles is the Coulomb potential, the parameters correspond to
hydrogen-like ion ($H^{+}$) of Kestner-Sinano\={g}lu
model.\cite{Kestner62} The exact results of all three cases can be
found elsewhere,\cite{Kestner62,Brosens98} which is easy to check
the validity of the current method. These models are widely used as
a benchmark for checking the usefulness of various methods for
\emph{sign problem}, see for examples
Ref.\cite{Lyubartsev,Newman91,Hall88,Miura01}

Our Metropolis MC scheme is preformed based on Eq. 19. At each step,
 $H_{p}$ are calculated to determine the rejection
and acceptance. $det\Lambda$ is calculated by a certain algorithm
with the computational cost scaled by $N^{3}$.\cite{nr} This kind of
numerical technique enables us to perform the fermionic simulations
with reasonable computational time. There are two basic types of
moves in current simulations: (1) Displacement move, where all the
coordinates for a single particle are displaced uniformly; (2)
Standard bisection moves.\cite{Ceperley95,Chakravarty} The MC
procedure used in this work is wildly used by others. One MC step is
defined as one application of each procedure. Ten million MC steps
of calculation were carried out for each temperature. For a few
cases, 100 million MC steps are made to check the ergodic problem.
The results agree with the short runs within the error bars. To
further check the ergodic problem, the simulations are carried out
by a few random generated starting configurations. All simulations
converge to the same results. The energy is calculated based on the
thermodynamic estimator.\cite{Ceperley95} The energies are well
converged at M/$\beta$=20, 22 and 5 for case 1, 2 and 3
respectively.

The calculated thermal energy is in good agreement with the exact
one for all three cases studied. In case 3, the exact energy 2.647
of Ref. \cite{Kestner62} has been almost accurately reproduced,
which is 2.652$\pm$0.003 in current simulations. Fig. 2 shows the
thermal energy per particle as a function of temperature for Case 1
and 2, the corresponding exact results are also shown in Fig. 2 with
lines. As can be seen from the figure, the overall temperature
dependence is well reproduced by current calculations. The
calculated thermal energies agree very well with the exact value at
low temperature. The slight deviation at high temperature is due to
the fact that the first stage of approximation will result in error
when the number of beads is too small, which is the case for high
temperature.

We have calculated the pair correlation function (PCF) between
beads, which is defined as,
\[g(r)=\langle
\frac{2}{MN(N-1)}\sum_{\nu}^{M}\sum_{i}^{N-1}\sum_{j>i}^{N}\delta(r-|\vec{r}_{i}^{(\nu)}-\vec{r}_{j}^{(\nu)}|)\rangle.\]
It is known that,\cite{Miura01} comparing with boson and Boltzmann
systems, the fermionic PCF has a hole around the origin, which
reflects the Pauli exclusion principle. In Fig. 3, we present PCF
for case 1 and 2 at temperature of 0.2. From this figure, we can see
that, the pair correlation function clearly represents the effect
due to the Pauli exclusion principle. The similar behaviors are
observed for other temperature and systems.

The average sign reflects the signal-to-noise ratio, which directly
affects calculation precision and computation time needed. The
average sign is defined as $Sign=(N_{+}-N_{-})/(N_{+}+N_{-})$, where
$N_{+}$ and $N_{-}$ are the total positive and negative
configuration respectively. The lower panel of Fig. 4 shows the
average sign of current simulations via temperature. It can be seen
that, $Sign$ decreases with the decrease of the temperature.
However, for the studied systems, even at lowest temperature
(T=0.1), the average sign is quite high (around 0.1).  We also
calculated the average sign via the number of particles at T=0.5 for
both case 1 and 2, which is shown in the upper panel of Fig. 4.
Similarly, $Sign$ also decreases with the increase of the number of
particles. Although we have not completely solved the \emph{sign
problem}, our approach does much improve the sign decay rate with
both temperature and number of particles. Direct using of Eq. 4,
$Sign$ is about 0.01 at temperature of 0.8 for case 1. And for
temperature lower than 0.8, the large sign fluctuation makes MC
simulation difficult to obtain any useful information. According to
the data shown in Fig. 4, the maximum number of particles, which can
be handled in current method, should be in order of ten. Considering
both spin-up and -down, the maximum number of particles can be
around twenty, which could be particularly useful for atom and
molecular systems.

\section{Discussion and Conclusion}

In this paper, we have introduced an approach to reduce the fermion
sign fluctuation in finite temperature PIMC simulations. By this
method, configurations,  which probably cause the sign fluctuation,
are pre-calculated within two stages of coarse-grained
approximations, while the rest are treated exactly. After two stages
of coarse-grained approximations, at least for two-particle system,
the \emph{sign problem} is solved completely. Since the exchange
matrix $A$ is replaced by a non-local one ($\Lambda$), the collapse
of fermion system into a boson one at low temperature has been
effectively avoided. The pilot calculation was performed on three
model systems: six independent particles in a three-dimensional
harmonic well, six interacting particles in a three-dimensional
harmonic well, and hydrogen-like ion ($H^{+}$) of
Kestner-Sinano\={g}lu model. The calculation shows that the current
approach not only dramatically drops the sign fluctuation, but also
gives an excellent description to real systems. Our method could be
particularly useful for atom and molecular systems. Although our
approach suffers from the \emph{sign problem} for large number of
particles, we believe that it provide an alternative thought on the
sign problem. We also believe that a similar approach can also be
helpful in other path integral methods. The current formula can be
easily extended to systems consisting of both spin-up and -down
fermions.(see for example, \cite{Oh98,Takahashi})

Our approximation breaks down for systems including particles more
than twenty (including both spin up and down particles). It would be
possible to generalize our method for problems of larger numbers of
fermions. Although we have used $\eta(r_{ij}^{(M)})=1$ through out
this paper, other values are also possible. For example, if
$\eta(r_{ij}^{(M)})=exp(-\alpha\frac{\beta}{2M}m\omega_{M}^{2}(r_{ij}^{(M)})^{2}$)
is chosen, the current method can be more flexible. For $\alpha=0$,
it is the case used in current work. For $\alpha=1$, $det\Lambda$
becomes the exact density matrix of free particles(see APPENDIX),
thus the sign problem can be avoided completely. In fact, with
$\alpha$ increasing from 0 to 1, the approximation becomes more and
more crude, but the negative parts become less and less. To further
improve the current method, a better form or value for
$\eta(r_{ij}^{(M)})$ could be found. It is actually the issue on
which we are working now.

\begin{acknowledgments}
I am very grateful to Prof. X. G. Gong and Prof. T. Xiang for
valuable discussions and encouragements. And thank Prof. Feng Zhou
for interesting discussions. I also would like to thank Guanwen
Zhang for reading the manuscript prior to publication and for
helpful suggestions. This work is supported by the National Natural
Science Foundation of China, Shanghai Project for the Basic
Research. The computation is performed in the Supercomputer Center
of Shanghai.
\end{acknowledgments}

{\bf APPENDIX}

 In this appendix, we will prove that the current
formula is exact for free particles. Since the second stage of
approximation is just a straightforward integration for free
particles, we only prove the formula of first stage is correct for
free particles. All Cartesian coordinates are equivalent for free
particles, for simplicity we only prove it in one Cartesian
direction, say, \emph{x}.

The partition function for free particles in one dimension has the
form,

\begin{equation}
Z=\frac{1}{N!}\int \prod_{i=1}^{N}
Dx_{i}^{(M)}\rho(\{x_{i}^{(M)}\},\{x_{i}^{(M)}\})
\label{eq:partation1}
\end{equation}

where $\rho(x_{i}^{(M)},x_{i}^{(M)})$ is the density matrix, of which
element with current formula reads,
\begin{equation}
\rho_{ij}(\{x_{i}^{M}\},\{x_{i}^{M}\})=C_{1D}\int\prod_{i=1}^{N}\prod_{\nu=1}^{M-1}Dx_{i}^{(\nu)}\Xi_{ij}^{(1D)}
exp(-\beta\sum_{i=1}^{N}\frac{1}{2}m\omega_{M}^{2}L_{ix}^{2})
\end{equation}
where $C_{1D}=(\frac{mM}{2\pi\beta\hbar^{2}})^{\frac{M}{2}}$, and
$\Xi_{ij}^{(1D)}$ is the one-dimensional counterpart of $\Xi_{ij}$,
which is
\begin{equation}
\Xi_{ij}^{(1D)}=\frac{\int
dx_{i}^{(1)}\bar{\Delta}(L_{ix},x_{i}^{(1)},x_{i}^{(M)})e^{-\frac{1}{2}\beta
m\omega^{2}_{M}((x_{i}^{(1)}-x_{j}^{(M)})^{2}-(x_{i}^{(1)}-x_{i}^{(M)})^{2}
)}}{\Upsilon_{1D}(L_{xi},x_{i}^{(M)})}
\end{equation}
\[\Upsilon_{1D}(L_{ix},x_{i}^{(M)})=\int dx_{i}^{(1)}\bar{\Delta}(L_{ix},x_{i}^{(1)},x_{i}^{(M)})\]

Eq. 22 is only relevant to $\{L_{ix}\}$ and  $\{x_{i}^{(1)}\}$, the
integration over $x_{i}^{(\nu)}$ ($\nu$=1,...M-1) can be replaced by
(M-1)-dimensional spherical polar coordinates, \emph{i.e},
integration over $L_{ix}$ multiplying
$\Upsilon_{1D}(L_{ix},x_{i}^{M})$, Eq. 22 becomes,

\begin{equation}
\rho_{ij}(\{x_{i}^{M}\},\{x_{i}^{M}\})\propto C_{1D}\int
dL_{xi}dx_{i}^{(1)}\bar{\Delta}(L_{ix},x_{i}^{(1)},x_{i}^{(M)})
\end{equation}
\[\times exp(-\frac{1}{2}\beta m\omega_{M}^{2}(L_{ix}^{2}+(x_{i}^{(1)}-x_{j}^{(M)})^{2}-(x_{i}^{(1)}-x_{i}^{(M)})^{2}))\]

Remembering for fixed $x_{i}^{(1)}$ and $x_{i}^{(M)}$, the minimum
value of $L_{ix}^{2}$ equals to
$(x_{i}^{(1)}-x_{i}^{(M)})^{2}\frac{M}{M-1}$. We first do the
integration over variable $\{L_{ix}\}$ by change of variables as
$L_{ix}^{'2}=L_{ix}^{2}-(x_{i}^{(1)}-x_{i}^{(M)})^{2}\frac{M}{M-1}$,
 $\rho_{ij}(\{x_{i}^{M}\},\{x_{i}^{M}\})$ becomes,

\begin{equation}
\rho_{ij}(\{x_{i}^{M}\},\{x_{i}^{M}\})\propto
C_{1D}C_{\bar{\Delta}}\frac{(\frac{M-4}{2})!}{(\frac{\beta
m\omega_{M}^{2}}{2})^{\frac{M-2}{2}}}\int dx_{i}^{(1)}
\end{equation}
\[ \times exp(-\frac{1}{2}\beta m\omega^{2}_{M}((x_{i}^{(1)}-x_{j}^{(M)})^{2}+\frac{1}{M-1}(x_{i}^{(1)}-x_{i}^{(M)})^{2})) \]
\[=C_{1D}C_{\bar{\Delta}}\frac{(\frac{M-4}{2})!}{(\frac{\beta m\omega_{M}^{2}}{2})^{\frac{M-2}{2}}}\sqrt{\frac{2(M-1)\pi}{M\beta m\omega_{M}^{2}}}exp(-\frac{1}{2M}\beta m\omega^{2}_{M}(x_{i}^{(M)}-x_{j}^{(M)})^{2}) \]
\[=C_M(\frac{m}{2\pi\beta\hbar^{2}})^{\frac{1}{2}}exp(-\frac{1}{2M}\beta m\omega^{2}_{M}(x_{i}^{(M)}-x_{j}^{(M)})^{2}), \]
where $C_{M}$ is an irrelevant constant.  The above express is the
exact formula for free particles. It needs to noted that, since we
only can get the relative value for $\bar{\Delta}$, we could not
obtain the absolute value of $C_{M}$. However the absolute value of
$C_{M}$ is irrelevant to our calculation, which is a function of
\emph{M} only. Our numerical test also shows that the calculated
element of density matrix based on the current formula is in
excellent agreement with the exact data. In fact, the current
formula must be exact for free particles, since all the
approximations become exact without potential part.

\begin{figure}[figure1]
\centering
\includegraphics[angle=-0,width=100mm]{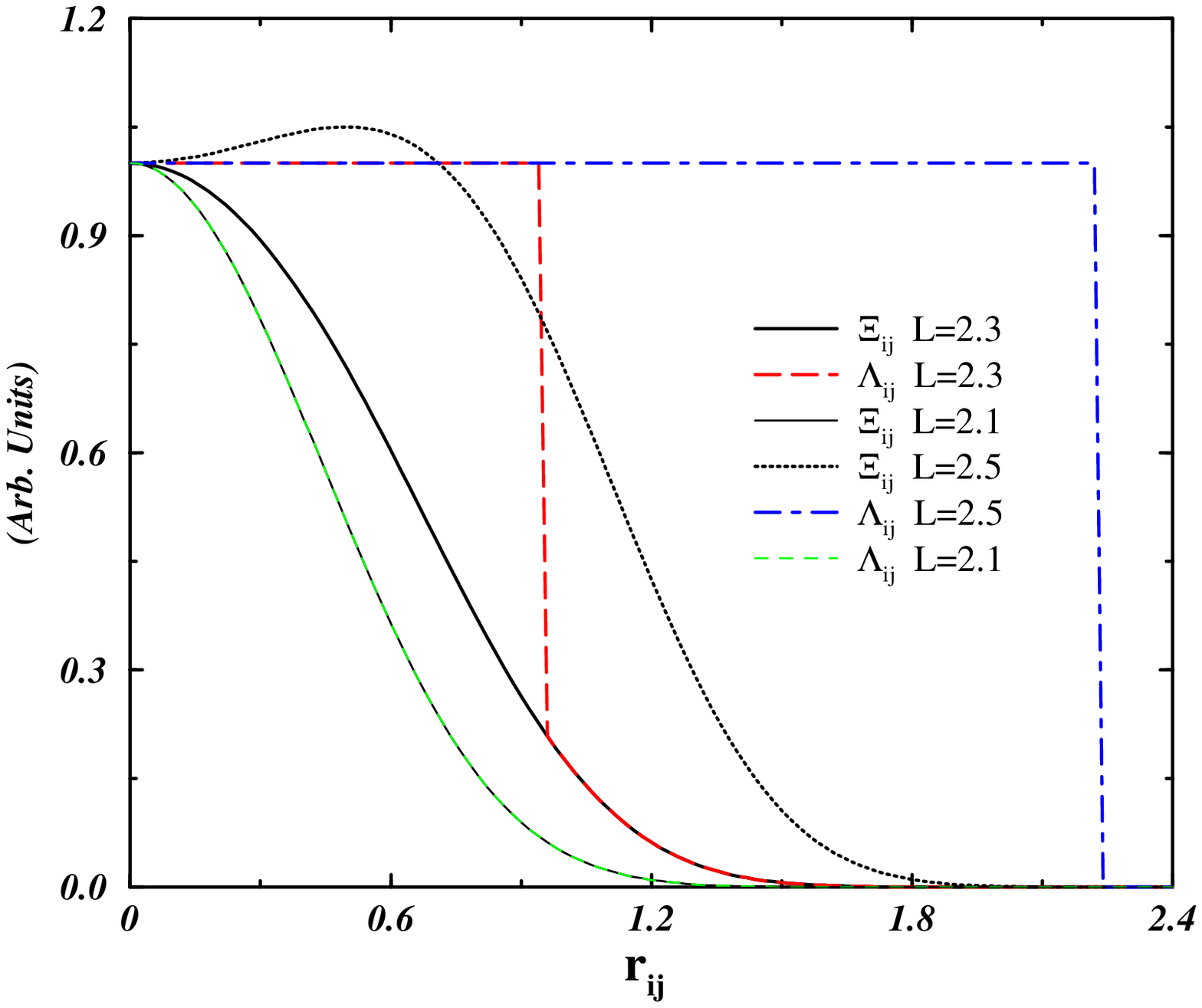}
\caption{(color online) The off-diagonal element of $\Lambda$ and
$\Xi$ ($\Lambda_{ij}$ and $\Xi_{ij}$) as a function of particle
separation ($r_{ij}$) for a few selected lengths of path ($L$) at
T=0.5. When the length of path is short, $\Lambda_{ij}$ and
$\Xi_{ij}$ are the same. As length being longer, $\Xi_{ij}$ can be
larger than one for short particle separation. In contrast,
$\Lambda_{ij}$ is not larger than one for any case.}
\end{figure}

\begin{figure}[figure2]
\centering
\includegraphics[angle=-0,width=100mm]{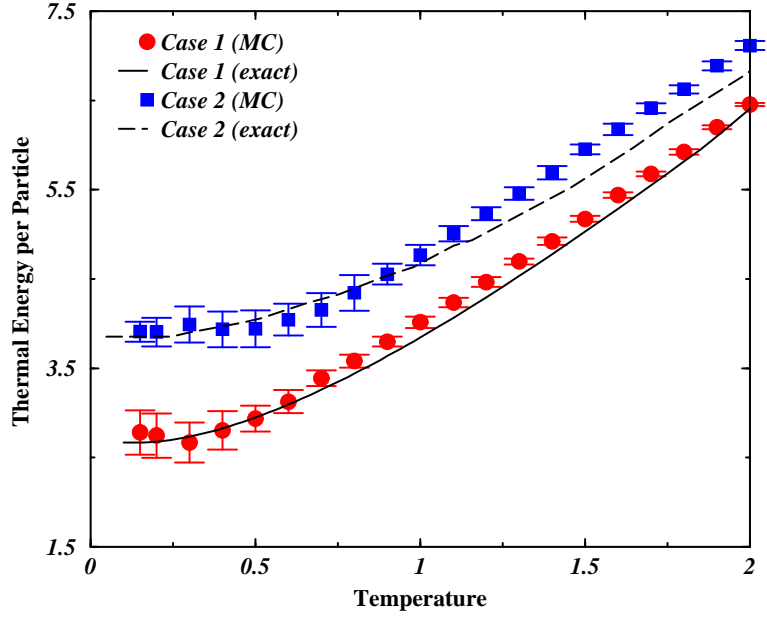}
\caption{(color online) The thermal energy of case 1 (Circle) and 2
(Squares) as a function of temperature calculated by PIMC. The solid
and dash lines are the exact energies of case 1 and 2 respectively.
The agreement is quite well.}
\end{figure}

\begin{figure}[figure3]
\centering
\includegraphics[angle=-90,width=100mm]{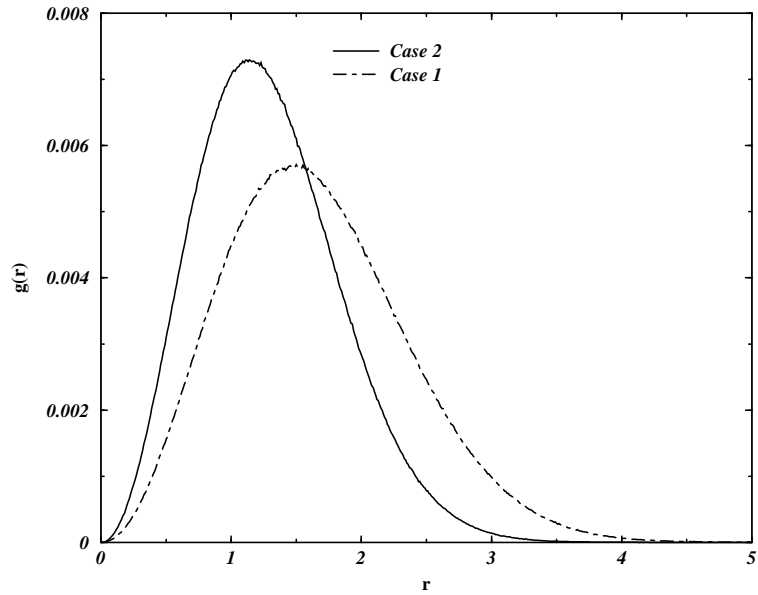}
\caption{The pair correlation functions between beads at T=0.2 for
both case 1 and 2. The hole around original is the reflection of
Pauli exclusion principle.}
\end{figure}

\begin{figure}[figure4]
\centering
\includegraphics[angle=-0,width=100mm]{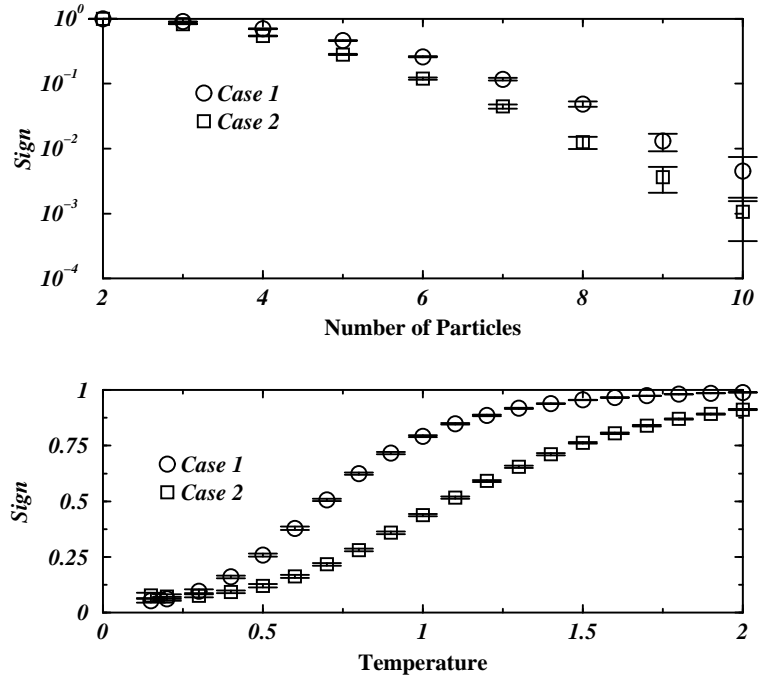}
\caption{The average sign of case 1 (circles) and 2 (squares)
 via temperature (lower panel) and number of particles (upper panel) as calculated
by our PIMC simulation. \emph{Sign} decreases with the decrease of
temperature, and the increase of number of particles.}
\end{figure}

\end{document}